\documentstyle[12pt,amssymb,amsmath]{article}
\newcommand{\p}{\partial}
\newcommand{\la}{\lambda}
\newcommand{\La}{\Lambda}

\begin{document}
\title{ {\bf B\"{a}cklund transformations for  high-order
constrained flows of the AKNS hierarchy:  canonicity and spectrality property}}
\author{ {\bf Yunbo Zeng\dag\hspace{1cm}  Huihui Dai\ddag \hspace{1cm} Jinping 
Song*} \\
    {\small {\it \dag
	Department of Mathematical Sciences, Tsinghua University,
	Beijing 100084, China}} \\
    {\small {\it \ddag
	Department of Mathematics, City University of Hong Kong,
	Kowloon, Hong Kong, China}}  \\  
    {\small {\it *
	Department of Mathematics, Henan University, Kaifeng 475001,
	Henan, China}}  } 
\date{}
\maketitle

\renewcommand{\theequation}{\arabic{section}.\arabic{equation}}

{\bf Abstract } 
New infinite number of one- and two-point B\"{a}cklund transformations (BTs) with 
explicit expressions are constructed for the high-order constrained flows of the 
AKNS hierarchy. It is shown that these BTs are canonical transformations including 
B\"{a}cklund parameter $\eta$ and a spectrality property holds with respect to 
$\eta$ and  the 'conjugated' variable $\mu$ for which the point $(\eta, \mu)$ 
belongs to the spectral curve. Also the formulas of m-times repeated Darboux 
transformations for the high-order constrained flows of the AKNS hierarchy are 
presented.

\section{Introduction}
\setcounter{equation}{0}
\hskip\parindent
B\"{a}cklund transformations (BTs) are an important aspect of the theory of
integrable systems \cite{a,1}. It is well-known that the BT's for
soliton equations are canonical transformations ( see, for emaple, \cite{fla,toda,
koda}). More recently there has been much interest in the
property of BTs for finite-dimensional integrable Hamiltonian systems 
\cite {2,3,4,5,6}. These BTs are defined as symplectic integrable maps which can 
be described explicitly and can be viewed as time discretization of particular 
flows of Liouville integrable systems. They are canonical transformations 
including B\"{a}cklund parameter $\eta$. For these BTs a spectrality property 
holds with respect to $\eta$ and  the 'conjugated' variable $\mu$ and the point 
$(\eta, \mu)$ or $(\eta, f(\mu))$ for some function $f(\mu)$ lies on the spectral 
curve.
An important application of the spectrality property of BTs is that to the problem 
of separation of variables. In fact, the sequence of B\"{a}cklund parameters 
$\eta_j$ together with the conjugate variables $\mu_j$ constitute the separation 
variables for the finite-dimensional integrable Hamiltonian systems \cite{2}.

We proceed to develop the ideas with some new BTs and study the problem of 
constructing one- and two-point B\"{a}cklund transformation for the high-order 
constrained flows of the soliton hierarchy \cite{7,8,9}. The Lax representation 
for the high-order constrained flows can always be deduced from the adjoint 
representation for the soliton hierarchy \cite{10,11}. Then, based on the results 
of Darboux transformations (DTs) for soliton hierarchy \cite{1,12,13,14,15}, we 
can find the DTs for the high-order constrained flows. By using the Lax 
representation, these DTs give rise to explicit one- and two-point BTs including 
one and two B\"{a}cklund parameters $\eta_i$, respectively. We  show  these BTs to 
be canonical transformations by presenting their generating functions. Then we 
show that these BTs possess the spectrality property with respect to $\eta$ and 
conjugate variable $\mu$ and the pairs $(\eta_i,\mu_i)$ belong to the spectral 
curve, namely satisfy the separation equations. Few example of this kind of BTs 
were presented in \cite{2,3,4,5,6}. This paper presents a way to find infinite 
number of BTs with the property described in \cite{2,3,4,5,6} by means of DTs for 
the high-order constrained flows of soliton hierarchy.
We will use the high-order constrained flows of the AKNS hierarchy to
illustrate the ideas.

In section 2, we briefly describe the high-order constrained flows of the AKNS 
hierarchy. In section 3 we first present three kinds of DTs for the constrained 
flows of the AKNS hierarchy. Then we find infinite number of new one-point and 
two-point BTs from the first and third kind of DTs, respectively, and show them to 
be canonical transformations and  possessing the spectrality property by using the 
first three high-order constrained flows as model in the section 3 and 4, 
respectively. Finally, the formula for m-times repeated DTs for the constrained 
flows are presented in section 5.

\section{High-order constrained flows of the AKNS hierarchy}
\setcounter{equation}{0}
\hskip\parindent
Let us briefly describe the high-order constrained flows of AKNS hierarchy. 
Consider the AKNS spectral problem \cite{Ablo}
\begin{equation}
\label{s1}
	\psi_x=U(u,\la)\psi\equiv \left( \begin{array}{cc}
	-\lambda & q \\ r&\lambda\end{array} \right)\psi,\qquad
\psi=\left( \begin{array}{c} 
	\psi_1 \\ \psi_2 \end{array} \right),\qquad
u=\left( \begin{array}{c} 
	q \\ r \end{array} \right),
\end{equation}
and the evolution of $\psi$
\begin{equation}
\label{s2}
		\psi_{t_n}=
	V^{(n)}(u, \lambda)\psi=
\sum_{i=0}^{n} \left( \begin{array}{cc}
	a_i & b_i \\ c_i & -a_i \end{array} \right)\lambda^{n-i}\psi,   
\end{equation}
where
$$a_0=-1,\quad b_0=c_0=a_1=0, \quad b_1=q,\quad c_1=r$$
$$a_2=\frac 12qr,\quad b_2=-\frac 12q_x,\quad c_2=\frac 12r_{x},...,$$
and in general
$$\left( \begin{array}{c} 
	c_{m+1} \\ b_{m+1} \end{array} \right)=L\left( \begin{array}{c} 
	c_{m} \\ b_{m} \end{array} \right)=L^m\left( \begin{array}{c} 
	r \\ q \end{array} \right), \quad
 a_{m,x}=qc_{m}-rb_{m},$$
$$L=\frac 12\left( \begin{array}{cc}
	D-2rD^{-1}q &2rD^{-1}r \\ -2qD^{-1}q&-D+2qD^{-1}r\end{array} \right),\quad 
D=\frac {\partial}{\partial x},\quad DD^{-1}=D^{-1}D=1.$$

Then the compatibility condition of Eqs. (\ref{s1}) and (\ref{s2})
gives rise to the AKNS hierarchy \cite{Ablo}
\begin{equation}
\label{s3}
	u_{t_n}=\left( \begin{array}{c} 
	q \\ r \end{array} \right)_{t_n}
=J\left( \begin{array}{c} 
	c_{n+1} \\ b_{n+1} \end{array} \right)
=J\frac{\delta H_{n+1}}{\delta u},
 \qquad n=0,1,\cdots,
\end{equation}
where 
$$H_n=\frac {2a_{n+1}}{n}, \qquad
J=\left( \begin{array}{cc}
	0 & -2 \\ 2&0\end{array} \right).$$
We have
$$\frac{\delta \lambda}{\delta q}=\psi_2^2,\qquad 
\frac{\delta \lambda}{\delta r}=-\psi_1^2.$$

The high-order constrained flows of the AKNS hierarchy consist of the equations 
obtained from the spectral problem (\ref{s1}) for $N$ distinct 
$\lambda_j$
and the restriction of the variational derivatives for conserved quantities $H_n$ 
and $\lambda_j$ \cite{7,11}
\begin{subequations}
\label{s4}
\begin{equation}
\label{s4a}
\Phi_{1,x}=-\Lambda\Phi_{1}+q\Phi_{2},\qquad
\Phi_{2,x}=r\Phi_{}+\Lambda\Phi_{2},\qquad
 \end{equation}
\begin{equation}
\label{s4b}
\frac{\delta H_{n+1}}{\delta u}-
	\alpha\sum_{j=1}^N 
	\frac{\delta \lambda_j}{\delta u}
=\left( \begin{array}{c} 
	c_{n+1} \\ b_{n+1} \end{array} \right)
-\alpha\left( \begin{array}{c} 
	<\Phi_2, \Phi_2> \\-<\Phi_1, \Phi_1> \end{array} \right)=0,
\end{equation}
\end{subequations}
where we have used $(\phi_{1j},\phi_{2j})^T$ to denote the solution of
(2.1) with $\la=\la_j, j=1,...,N$ and $ \Phi_{i}=(\phi_{i1},...,\phi_{iN})^T, 
i=1,2,
\Lambda=\mbox{diag} (\lambda_1,...,\lambda_N), <.,.> $ denotes the inner product. 
The Lax representation for the constrained flow (\ref{s4}) is given by 
\cite{10,11}
\begin{equation}
\label{s5}
M^{(n)}_x=[U, M^{(n)}],
\end{equation}
with Lax matrix $M^{(n)}$
\begin{equation}
\label{s6}
 M^{(n)}(u, \Phi_1, \Phi_2, \la)=\left( \begin{array}{cc}
	A^{(n)} & B^{(n)} \\C^{(n)} & -A^{(n)}\end{array} \right) 
=V^{(n)}+M_0
\end{equation}
 $$M_0=\alpha\sum_{j=1}^{N} \frac{1}{\lambda-\lambda_j}
	\left( \begin{array}{cc}
	\phi_{1j}\phi_{2j} & -{\phi_{1j}}^2 \\ 
	{\phi_{2j}}^2 & -\phi_{1j}\phi_{2j} 
	\end{array} \right),$$
and the Lax pair for (\ref{s4})
\begin{equation}
\label{s7}
	\psi_x=
	U(u,\lambda)\psi,
\end{equation}
\begin{equation}
\label{s8}
                M^{(n)}(u, \Phi_1, \Phi_2, \lambda) \psi=\mu\psi.
\end{equation}
The spectral curve $\Gamma$,
$$\Gamma:\quad det(M^{(n)}(u, \Phi_1, \Phi_2,\lambda)-\mu)=0$$
is
\begin{equation}
\label{s9}
\mu^2=(A^{(n)})^2(\lambda)+B^{(n)}(\lambda)C^{(n)}(\lambda).
\end{equation}

We present the first three high-order constrained flows as follows.

(1) For $n=0, \alpha=\frac 12$, (\ref{s4b}) gives an explicit constraint
\begin{equation}
\label{s10}
	q=-\frac 12<\Phi_{1},\Phi_{1}>, \qquad
r=\frac 12<\Phi_{2},\Phi_{2}>.
\end{equation}
Then (\ref{s4a}) becomes a finite-dimensional integrable Hamiltonian system 
(FDIHS)
 \begin{equation}
\label{s11}
\Phi_{1,x}=\frac {\p \widetilde H_0}{\p\Phi_2},\qquad
\Phi_{2,x}=-\frac {\p \widetilde H_0}{\p\Phi_1},
\end{equation}
$$\widetilde H_0=-<\La\Phi_1,\Phi_{2}>-\frac 
14<\Phi_1,\Phi_{1}><\Phi_2,\Phi_{2}>,$$
with Lax matrix $M^{(0)}$
$$A^{(0)}=-1+ \frac 12\sum_{j=1}^N \frac{1}{\lambda-\lambda_j}
	\phi_{1j}\phi_{2j}, \qquad
B^{(0)}=-\frac 12\sum_{j=1}^N \frac{1}{\lambda-\lambda_j}
	\phi^2_{1j},$$
 \begin{equation}
\label{s12}
C^{(0)}= \frac 12\sum_{j=1}^N \frac{1}{\lambda-\lambda_j}
	\phi^2_{2j}.
\end{equation}
The spectral curve $\Gamma$ is a hyperelliptic, genus $N-1$ curve
\begin{equation}
\label{s13}
\mu^2=1+\sum_{j=1}^N \frac{P_j}{\lambda-\lambda_j},
\end{equation}
with
$$P_j=-\phi_{1j}\phi_{2j}+\frac 12\sum_{k\neq j} 
\frac{1}{\lambda_j-\lambda_k}(\phi_{1j}\phi_{2j} 
\phi_{1k}\phi_{2k}-\phi^2_{1k}\phi^2_{2j}),\quad j=1,...,N.$$
$P_1,...,P_N$ are $N$ independent integrals of motion in involution for FDIHS 
(\ref{s11}).

(2) For $n=1, \alpha=-\frac 14$, (\ref{s4}) can be written as  a FDIHS
 \begin{equation}
\label{s14}
Q_{x}=\frac {\p \widetilde H_1}{\p P},\qquad
P_{x}=-\frac {\p \widetilde H_1}{\p Q},
\end{equation}
with
$$\widetilde H_1=-<\La\Phi_1,\Phi_{2}>-\frac 12 r<\Phi_1,\Phi_{1}>+\frac 12 
q<\Phi_2,\Phi_{2}>,$$
$$Q=(\phi_{11},...,\phi_{1N},q)^T,\qquad 
P=(\phi_{21},...,\phi_{2N},r)^T,$$
and Lax matrix $M^{(1)}$
$$A^{(1)}=-\la- \frac 14\sum_{j=1}^N \frac{1}{\lambda-\lambda_j}
	\phi_{1j}\phi_{2j}, \qquad
B^{(1)}=q+\frac 14\sum_{j=1}^N \frac{1}{\lambda-\lambda_j}
	\phi^2_{1j},$$
 \begin{equation}
\label{s15}
C^{(1)}=r- \frac 14\sum_{j=1}^N \frac{1}{\lambda-\lambda_j}
	\phi^2_{2j}.
\end{equation}
The spectral curve $\Gamma$ is
\begin{equation}
\label{s16}
\mu^2=\la^2+P_0+\sum_{j=1}^N \frac{P_j}{\lambda-\lambda_j},
\end{equation}
with
$$P_0=\frac 12<\Phi_{1}\Phi_{2}>+qr,$$
$$P_j=\frac 14(2\la_j\phi_{1j}\phi_{2j}+r\phi^2_{1j}-q\phi^2_{2j})+\frac 
18\sum_{k\neq j} 
\frac{1}{\lambda_j-\lambda_k}(\phi_{1j}\phi_{2j}\phi_{1k}\phi_{2k}-\phi^2_{1k}\phi
^2_{2j}).$$
$P_0,...,P_N$ are $N+1$ independent integrals of motion in involution for FDIHS 
(\ref{s14}).

(3) For $n=2, \alpha=\frac 12$, by introducing the following Jacobi-Ostrogradsky 
coordinates
$$Q=(\phi_{11},...,\phi_{1N},q_1,q_2)^T,\qquad 
P=(\phi_{21},...,\phi_{2N},r_1,r_2)^T,$$
$$q_1=q,\qquad q_2=r,\qquad 
p_1=-\frac 14r_x,\qquad p_2=-\frac 14 q_x,$$
(\ref{s4}) can be transformed into a FDIHS
 \begin{equation}
\label{s17}
Q_{x}=\frac {\p \widetilde H_2}{\p P},\qquad
P_{x}=-\frac {\p \widetilde H_2}{\p Q},
\end{equation}
with
$$\widetilde H_2=-<\La\Phi_1,\Phi_{2}>-\frac 12q_2<\Phi_1,\Phi_{1}>+\frac 
12q_1<\Phi_2,\Phi_{2}>+\frac 14q_1^2q_2^2-4p_1p_2,$$
and Lax matrix $M^{(2)}$
$$A^{(2)}=-\la^2+\frac 12q_1q_2+ \frac 12\sum_{j=1}^N \frac{1}{\lambda-\lambda_j}
	\phi_{1j}\phi_{2j}, \qquad
B^{(2)}=q_1\la+2p_2-\frac 12\sum_{j=1}^N \frac{1}{\lambda-\lambda_j}
	\phi^2_{1j},$$
 \begin{equation}
\label{s18}
C^{(2)}=q_2\la-2p_1+\frac 12 \sum_{j=1}^N \frac{1}{\lambda-\lambda_j}
	\phi^2_{2j}.
\end{equation}
The spectral curve $\Gamma$ is
\begin{equation}
\label{s19}
\mu^2=\la^4+P_0\la+P_{N+1}+\sum_{j=1}^N \frac{P_j}{\lambda-\lambda_j},
\end{equation}
with
$$P_0=-<\Phi_{1}\Phi_{2}>-2q_1p_1+2p_2q_2, \qquad P_{N+1}=\widetilde H_2,$$
$$P_j=\frac 
12(-2\la^2_j\phi_{1j}\phi_{2j}-\la_jq_2\phi^2_{1j}+\la_jq_1\phi^2_{2j}+q_1q_2\phi_
{1j}\phi_{2j})+p_1\phi^2_{1j}+p_2\phi^2_{2j}$$
$$+\frac 12\sum_{k\neq j} 
\frac{1}{\lambda_j-\lambda_k}(\phi_{1j}\phi_{2j}\phi_{1k}\phi_{2k}-\phi^2_{1k}\phi
^2_{2j}),\qquad  j=1,...,N.$$
$P_0,...,P_{N+1}$ are $N+2$ independent integrals of motion in involution for 
FDIHS (\ref{s17}).

\section{One-point BTs for high-order constrained flows of the AKNS hierarchy}
\setcounter{equation}{0}
\hskip\parindent
We first briefly review the Darboux transformations (DTs) for the AKNS hierarchy. 
Suppose that a gauge transformation
\begin{equation}
\label{h1}
	\bar\psi=T\psi
\end{equation}
transforms (\ref{s1}) and (\ref{s2}) into
\begin{equation}
\label{h2}
	\bar\psi_x=\overline U(\bar u,\la)\bar\psi,
\end{equation}
\begin{equation}
\label{h3}
	 \bar\psi_{t_n}=
\overline V^{(n)}(\bar u, \lambda)\bar\psi.
\end{equation}
Let $\psi(x,\eta_i)$ be a solution of (\ref{s1}) and (\ref{s2}) with $\la=\eta_i, 
i=1,2, \eta_i\neq\la_j.$ It is known \cite{1,12,13,14} that
there are the following three kinds of the DTs for the AKNS hierarchy.

(1) The first DT for the AKNS hierarchy is given by
\begin{equation}
\label{h4}
	T_1= \left( \begin{array}{cc}
	\lambda-\eta_1+\frac 12 qf_1 &-\frac 12q \\-f_1&1\end{array} \right), 
\qquad f_i=\frac {\psi_2(x,\eta_i)}{\psi_1(x,\eta_i)},
\end{equation}
and
 \begin{equation}
\label{h5}
\bar q=-\frac 12q_x-\eta_1q+\frac 12q^2f_1, \qquad \bar r=2f_1,
\end{equation}
namely under the transformation (\ref{h1}) with (\ref{h4}) and (\ref{h5}), 
$\overline U$ and $\overline V^{(n)}$ are of the same form as $U$ and $V^{(n)}$ 
except for replacing $q$ and $r$ by $\bar q$ and $\bar r$. So (\ref{h5}) presents 
the relationship between two solutions $(q,r)$ and $(\bar q,\bar r)$ of the 
equation (\ref{s3}).

(2) The second DT for the AKNS hierarchy is given by
\begin{equation}
\label{h6}
	T_2= \left( \begin{array}{cc}
	1 &-f_2\\\frac 12r&\lambda-\eta_2-\frac 12 rf_2\end{array} \right), 
\end{equation}
and
 \begin{equation}
\label{h7}
\bar q=-2f_2, \qquad \bar r=\frac 12r_x-\eta_2r-\frac 12r^2f_2.
\end{equation}
(3) The third DT for the AKNS hierarchy is given by
\begin{equation}
\label{h8}
	T_3= \left( \begin{array}{cc} \lambda-\eta_1+m_2
	 &-m_1\\m_3&\lambda-\eta_2-m_2\end{array} \right), 
\end{equation}
and
 \begin{equation}
\label{h9}
\bar q=q-2m_1, \qquad \bar r=r-2m_3,
\end{equation}
with
$$m_1=\frac {(\eta_2-\eta_1)\psi_1(\eta_1)\psi_1(\eta_2)}{\triangle}, \qquad
m_2=\frac {(\eta_2-\eta_1)\psi_1(\eta_2)\psi_2(\eta_1)}{\triangle},$$
\begin{equation}
\label{hh9}
m_3=\frac {(\eta_2-\eta_1)\psi_2(\eta_1)\psi_2(\eta_2)}{\triangle}, \qquad
{\triangle}=\psi_1(\eta_1)\psi_2(\eta_2)-\psi_2(\eta_1)\psi_1(\eta_2).
\end{equation}

We now consider the DTs for high-order constrained flows (\ref{s4}). 
 Suppose that the gauge transformation (\ref{h1}) and accordingly
\begin{equation}
\label{hh}
\left( \begin{array}{c} 
      \bar  \phi_{1j} \\ \bar\phi_{2j} \end{array} \right)
=\beta_j T\left( \begin{array}{c} 
       \phi_{1j} \\ \phi_{2j} \end{array} \right),
\end{equation}                                             
transforms (\ref{s7}) and (\ref{s8}) into
\begin{equation}
\label{h10}
	\bar\psi_x=\overline U(\bar u,\la)\bar\psi,
\end{equation}
\begin{equation}
\label{h11} 
         \overline M^{(n)}(\bar u, \bar \Phi_1, \bar \Phi_2,\lambda)\bar\psi
         =\mu\bar\psi,
\end{equation}
where $\overline U$ and $\overline M^{(n)}$ satisfy
\begin{equation}
\label{h12}
	T_x=\overline U(\bar u,\la)T-TU(u,\la),
\end{equation}
\begin{equation}
\label{h13}
         \overline M^{(n)}(\bar u,\bar \Phi_1, \bar \Phi_2,\la)T
         =TM^{(n)}(u, \Phi_1, \Phi_2,\la).
\end{equation}
 Motivated by the first DT for the AKNS hierarchy, let $\psi(x,\eta_i)$ be a 
solution of (\ref{s7}) and (\ref{s8}) with
 $\la=\eta_i, \mu=\mu_i, i=1,2, \eta_i\neq\la_j.$
 We find that the first DT for the constrained flows (\ref{s4})
 consists of (\ref{h1}), (\ref{h4}), (\ref{h5}) and  (\ref{hh}) with
 $\beta_j=\frac {1}{\sqrt{\la_j-\eta_1}}$, namely
\begin{subequations}
\label{h14}
\begin{equation}
\label{h14a}
\bar\phi_{1j}=\sqrt{\la_j-\eta_1}\phi_{1j}
-\frac {1}{2\sqrt{\la_j-\eta_1}}q(\phi_{2j}-f_1\phi_{1j}),
\end{equation}
\begin{equation}
\label{h14b}
\bar\phi_{2j}=\frac {1}{\sqrt{\la_j-\eta_1}}(\phi_{2j}-f_1\phi_{1j}),
\qquad j=1,...,N.
\end{equation}
\end{subequations}
In fact, based on the results of the DTs for the AKNS hierarchy,
it can be shown by a similar  way in \cite{ze,15} that under the
transformation
(\ref{h1}), (\ref{h4}), (\ref{h5}) and (\ref{h14}), $\overline U$
and $\overline M^{(n)}$ permit  the same form as $U$ and $M^{(n)}$
except for replacing $q,r,\phi_{1j},$ $\phi_{2j}$ by $\bar q,\bar r,$
$\bar\phi_{1j},\bar\phi_{2j}$, namely, we have
\begin{equation}
\label{h12'}
        T_x=U(\bar u,\la)T-TU(u,\la),
\end{equation}
\begin{equation}
\label{h13'}
         M^{(n)}(\bar u,\bar\Phi_1, \bar\Phi_2, \la)T=TM^{(n)}(u,
         \Phi_1, \Phi_2, \la).
\end{equation}
The equalities (\ref{h12'}) and (\ref{h13'}) ensure that (2.7) and
(2.8) are invarint under the  transformation (\ref{h1}), (\ref{h4}),
(\ref{h5}) and (\ref{h14}). This guarantees that
the relationship between
$q,r,\phi_{1j},$ $\phi_{2j}$ and $\bar q,\bar r,$
$\bar\phi_{1j},\bar\phi_{2j}$ obtained from (\ref{h12'}) and (\ref{h13'}) is
just the one between two solutions of the constrained flows  (\ref{s4}).
In fact, (\ref{h12'}) and (\ref{h13'})  give rises to  (\ref{h5}) and
(\ref{h14}) which present the relationship
between two solutions of the constrained flows  (\ref{s4}).

It follows from (\ref{s8})
\begin{equation}
\label{h15}
f_i=\frac {\mu_i-A^{(n)}(\eta_i)}{B^{(n)}(\eta_i)}
=\frac {C^{(n)}(\eta_i)}{\mu_i+A^{(n)}(\eta_i)},\qquad i=1,2.
\end{equation}
By substituting (\ref{h15}) into (\ref{h5}) and (\ref{h14}), we obtain infinite 
number ($n=0,1,...$) of the first explicit one-point BT $B_{\eta_1}$ 
for the constrained flows (\ref{s4}) as follows
\begin{subequations}
\label{h16}
\begin{equation}
\label{h16a}
\bar q=-\frac 12q_x-\eta_1q+\frac 12q^2\frac 
{\mu_1-A^{(n)}(\eta_1)}{B^{(n)}(\eta_1)}
, \qquad \bar r=2\frac {\mu_1-A^{(n)}(\eta_1)}{B^{(n)}(\eta_1)},
\end{equation}
\begin{equation}
\label{h16b}
\bar\phi_{1j}=\sqrt{\la_j-\eta_1}\phi_{1j}
-\frac {q}{2\sqrt{\la_j-\eta_1}}(\phi_{2j}-\frac 
{\mu_1-A^{(n)}(\eta_1)}{B^{(n)}(\eta_1)}
\phi_{1j}),
\end{equation}
\begin{equation}
\label{h16c}
\bar\phi_{2j}=\frac {1}{\sqrt{\la_j-\eta_1}}(\phi_{2j}-\frac 
{\mu_1-A^{(n)}(\eta_1)}{B^{(n)}(\eta_1)}
\phi_{1j}).
\end{equation}
\end{subequations}

It is found from (\ref{h4}) and (\ref{h13'})
\begin{subequations}
\label{h17}
\begin{equation}
\label{h17a}
(\la-\eta_1)\overline 
A^{(n)}(\la)=(\la-\eta_1+qf_1)A^{(n)}(\la)+f_1(\la-\eta_1+\frac 12 
qf_1)B^{(n)}(\la)
-\frac 12q C^{(n)}(\la),
\end{equation}
\begin{equation}
\label{h17b}
(\la-\eta_1)\overline B^{(n)}(\la)=q(\la-\eta_1+\frac 12 
qf_1)A^{(n)}(\la)+(\la-\eta_1+\frac 12 qf_1)^2 B^{(n)}(\la)
-\frac 14q^2 C^{(n)}(\la),
\end{equation}
\begin{equation}
\label{h17c}
(\la-\eta_1)\overline C^{(n)}(\la)=-2f_1A^{(n)}(\la)-
f_1^2B^{(n)}(\la)+C^{(n)}(\la),
\end{equation}
\end{subequations}
which, as we mentioned above, present the relationship
between two solutions of the constrained flows  (\ref{s4}).

Using the first three constrained flows as model, we now show
the BTs (\ref{h16}) to be  canonical transformations by presenting
their generating functions and check spectrality property  with respect to the 
B\"{a}cklund parameter $\eta$ and  the 'conjugated' variable $\mu$ with the point 
$(\eta,\mu)$ belonging to the spectral curve (\ref{s9}).

(1) For the first constrained flow, the FDIHS (\ref{s11}), using (\ref{s12})
and comparing the
coefficients of $\la^0$ in (\ref{h17c}), one gets
\begin{equation}
\label{h18}
	f_1=\frac 14<\overline \Phi_2,\overline \Phi_2>.
\end{equation}
Then we have from (\ref{h14})
\begin{subequations}
\begin{equation}
\label{h19a}
	\phi_{2j}=\sqrt{\la_j-\eta_1} \bar\phi_{2j}
+\frac 14<\overline \Phi_2,\overline \Phi_2>\phi_{1j}=\frac {\p S^{(0)}}{\p 
\phi_{1j}},
\end{equation}
\begin{equation}
\label{h19b}
\overline\phi_{1j}=\sqrt{\la_j-\eta_1}\phi_{1j}
+\frac 14<\Phi_1,\Phi_1>\overline\phi_{2j}=\frac {\p S^{(0)}}{\p 
\overline\phi_{2j}},
\end{equation}
\end{subequations}
where the generating function $S^{(0)}$ for the canonical transformation 
(\ref{h16}) is given by
\begin{equation}
\label{h20}
S^{(0)}=\frac 18<\Phi_1,\Phi_1><\overline\Phi_{2},\overline\Phi_{2}>
+\sum_{j=1}^N\sqrt{\la_j-\eta_1}\phi_{1j}\overline\phi_{2j}-\eta_1.
\end{equation}
Furthermore, it is found from (\ref{s12}) and (\ref{h14})
$$\frac {\p S^{(0)}}{\p \eta_1}=-1-\frac 12\sum_{j=1}^N
\frac 1{\sqrt{\la_j-\eta_1}}\phi_{1j}\overline\phi_{2j}$$
\begin{equation}
\label{h21}
=-1-\frac 12\sum_{j=1}^N\frac 1{\sqrt{\la_j-\eta_1}}\phi_{1j}\frac 
1{\sqrt{\la_j-\eta_1}}[\phi_{2j}-f_1\phi_{1j}]
=A^{(0)}(\eta_1)+f_1B^{(0)}(\eta_1)=\mu_1,
\end{equation}
which implies that $(\eta_1,\mu_1)$  satisfies the spectrality property.
Consider the composition $B_{\eta_1...\eta_N}=B_{\eta_1}\circ...\circ  B_{\eta_N}$ 
of the B\"{a}cklund transformation $B_{\eta_i}$. Then the corresponding generating 
function $S^{(0)}_{\eta_1...\eta_N}$ becomes the generating function of the 
canonical transformation from $(\Phi_1,\Phi_2)$ to $(\eta, \mu)$ given by the 
equations
$$\phi_{2j}=\frac {\p S^{(0)}_{\eta_1...\eta_N}}{\p \phi_{1j}},\qquad
\mu_{j}=\frac {\p S^{(0)}_{\eta_1...\eta_N}}{\p \eta_{j}}.$$
The points $(\eta_i,\mu_i)$ satisfy the separation equations given by the spectral 
curve (\ref{s13})
$$\mu_i^2=1+\sum_{j=1}^N \frac{P_j}{\eta_i-\lambda_j}, \qquad i=1,...,N.$$

(2) For the second constrained flow, the  FDIHS (\ref{s14}),
using (\ref{s15}) and comparing the coefficients of $\la, \la^0$ in
(\ref{h17c}) and coefficient of $\la$ in (\ref{h17b}), one gets $f_1=\frac 12\bar 
r$ and
\begin{equation}
\label{h22c}
r=-\frac 14<\overline\Phi_2,\overline\Phi_2>-\eta_1\bar r+\frac 14q\bar r^2=\frac 
{\p S^{(1)}}{\p q},
\end{equation}
\begin{equation}
\label{h22d}
\bar q=\frac 14<\Phi_1,\Phi_1>-\eta_1q +\frac 14q^2\bar r=\frac {\p S^{(1)}}{\p 
\bar r},
\end{equation}
then using (\ref{h14})
\begin{subequations}
\label{h22}
\begin{equation}
\label{h22a}
	\phi_{2j}=\sqrt{\la_j-\eta_1}\bar\phi_{2j}
+\frac 12 \bar r\phi_{1j}=\frac {\p S^{(1)}}{\p \phi_{1j}},
\end{equation}
\begin{equation}
\label{h22b}
\overline\phi_{1j}=\sqrt{\la_j-\eta_1}\phi_{1j}
-\frac 12 q\overline\phi_{2j}=\frac {\p S^{(1)}}{\p \overline\phi_{2j}},
\end{equation}
\end{subequations}
where the generating function $S^{(1)}$ for the canonical transformation 
(\ref{h16}) is given by
\begin{equation}
\label{h23}
S^{(1)}=\frac 14\bar r<\Phi_1,\Phi_1>-\frac 14 
q<\overline\Phi_{2},\overline\Phi_{2}>-\eta_1q\bar r+\frac 18q^2\bar r^2
+\sum_{j=1}^N\sqrt{\la_j-\eta_1}\phi_{1j}\overline\phi_{2j}+\eta_1^2.
\end{equation}
Furthermore, it is easy to check the spectrality property by means of  (\ref{h14}) 
and (\ref{s15})
$$\frac {\p S^{(1)}}{\p \eta_1}=-\frac 12\sum_{j=1}^N\frac 
1{\sqrt{\la_j-\eta_1}}\phi_{1j}\overline\phi_{2j}-q\bar r +2\eta_1$$
\begin{equation}
\label{h24}
=-2[A^{(1)}(\eta_1)+f_1B^{(1)}(\eta_1)]=-2\mu_1.
\end{equation}
The point $(\eta_1,\mu_1)$ satisfies the separation equation given by the spectral 
curve (\ref{s16})
$$\mu_1^2=\eta_1^2+P_0+\sum_{j=1}^N \frac{P_j}{\eta_1-\lambda_j}.$$

(3) For the third constrained flow, the  FDIHS (\ref{s17}), using
(\ref{s18}), (\ref{h14}) and (\ref{h17}) in the same way as for (\ref{h22}), one 
gets
$f_1=\frac 12\bar q_2$ and 
\begin{subequations}
\begin{equation}
\label{h25a}
	\phi_{2j}=\sqrt{\la_j-\eta_1}\overline\phi_{2j}
+\frac 12 \bar q_2\phi_{1j}=\frac {\p S^{(2)}}{\p \phi_{1j}},
\end{equation}
\begin{equation}
\label{h25b}
\overline\phi_{1j}=\sqrt{\la_j-\eta_1}\phi_{1j}
-\frac 12 q_1\overline\phi_{2j}=\frac {\p S^{(2)}}{\p \overline\phi_{2j}},
\end{equation}
\begin{equation}
\label{h25c}
\bar q_1=\frac 14q_1^2\bar q_2-\eta_1 q_1+2p_2=\frac {\p S^{(2)}}{\p \bar p_1}
\end{equation}
\begin{equation}
\label{h25d}
q_2=\frac 14q_1\bar q_2^2-\eta_1 \bar q_2-2\bar p_1=-\frac {\p S^{(2)}}{\p  p_2}
\end{equation}
then $p_1=-\frac 14 q_{2x}$ and $\bar p_2=-\frac 14\bar q_{1x}$ lead
\begin{equation}
\label{h25e}
p_1=-\frac 14<\overline\Phi_2,\overline\Phi_2>-\eta_1\bar p_1 +\frac 12q_1\bar 
q_2\bar p_1-\frac 1{16}q_1^2\bar q_2^3+\frac 14q_1\bar q_2^2\eta_1-\frac 14\bar 
q_2^2 p_2=\frac {\p S^{(2)}}{\p q_1},
\end{equation}
\begin{equation}
\label{h25f}
\bar p_2=-\frac 14<\Phi_1,\Phi_1>-\eta_1 p_2 +\frac 12q_1\bar q_2 p_2+\frac 
1{16}q_1^3\bar q_2^2-\frac 14q_1^2\bar q_2\eta_1-\frac 14 q_1^2 \bar p_1=-\frac 
{\p S^{(2)}}{\p \bar q_2},
\end{equation}
\end{subequations}
where the generating function $S^{(2)}$ for the canonical transformation 
(\ref{h16}) is given by
$$S^{(2)}=\frac 14\bar q_2<\Phi_1,\Phi_1>-\frac 14 
q_1<\overline\Phi_{2},\overline\Phi_{2}>-\eta_1q_1\bar p_1
+\eta_1 \bar q_2p_2-\frac 1{4}q_1\bar q_2^2p_2$$
\begin{equation}
\label{h26}
+2\bar p_1p_2+\frac 14q_1^2\bar q_2\bar p_1+\frac 18 q_1^2 \bar q_2^2\eta_1
-\frac 1{48}q^3\bar 
q_2^3+\sum_{j=1}^N\sqrt{\la_j-\eta_1}\phi_{1j}\overline\phi_{2j}-\frac 13 
\eta_1^3.
\end{equation}
Then it is easy to check the spectrality
$$\frac {\p S^{(2)}}{\p \eta_1}=\frac 12\sum_{j=1}^N\frac 
{\phi_{1j}}{\la_j-\eta_1}[\phi_{2j}-\frac 12 \bar q_2\phi_{1j}]+\bar q_2p_2$$
\begin{equation}
\label{h27}
+\frac 12q_1[q_2-\frac 1{4}q_1\bar q_2^2+\bar q_2\eta_1]+\frac 18 q_1^2 \bar q_2^2
=A^{(2)}(\eta_1)+f_1B^{(2)}(\eta_1)=\mu_1.
\end{equation}
The point $(\eta_1,\mu_1)$ satisfies the separation equation given by the spectral 
curve (\ref{s19})
$$\mu_1^2=\eta_1^4+P_0\eta_1+P_{N+1}+\sum_{j=1}^N \frac{P_j}{\eta_1-\lambda_j}.$$

In the exactly same way, we can find second one-point BTs for the constrained 
flows (\ref{s4}) according to the second DTs for the AKNS hierarchy. By 
composition of these two BTs, we can find two-point BTs for the constrained flows 
(\ref{s4}). Since these two-point BTs are quite complicate, we will present 
another two-point BTs for the constrained flows in the next section.

\section{Two-point BTs for high-order constrained flows of the AKNS hierarchy}
\setcounter{equation}{0}
\hskip\parindent

Let 
$\psi(x,\eta_i)$ be a solution of (\ref{s7}) and (\ref{s8}) with $\la=\eta_i, 
\mu=\mu_i, i=1,2, \eta_i\neq\la_j.$  Motivated by the third DT for the AKNS 
hierarchy, we obtain the third DT for the constrained flows (\ref{s4}) consisting 
of (\ref{h1}), 
(\ref{h8}) and 
 \begin{subequations}
\label{g1}
\begin{equation}
\label{g1a}
\bar q=q-2m_1, \qquad \bar r=r-2m_3,
\end{equation}
\begin{equation}
\label{g1b}
\bar\phi_{1j}=\frac 
1{\sqrt{(\la_j-\eta_1)(\la_j-\eta_2)}}[(\la_j-\eta_1+m_2)\phi_{1j}-m_1\phi_{2j}],
\end{equation}
\begin{equation}
\label{g1c}
	\bar\phi_{2j}=\frac 
1{\sqrt{(\la_j-\eta_1)(\la_j-\eta_2)}}[m_3\phi_{1j}+(\la_j-\eta_2-m_2)\phi_{2j}].
\end{equation}
\end{subequations}
It follows from (\ref{hh9}) and (\ref{h15})
\begin{subequations}
\label{g2}
\begin{equation}
\label{g2a}
m_1=\frac {(\eta_2-\eta_1)B^{(n)}(\eta_1)B^{(n)}(\eta_2)}
{(\mu_2-A^{(n)}(\eta_2))B^{(n)}(\eta_1)-(\mu_1-A^{(n)}(\eta_1))B^{(n)}(\eta_2)},
\end{equation}
\begin{equation}
\label{g2b}
m_3=\frac {(\eta_2-\eta_1)(\mu_1-A^{(n)}(\eta_1))(\mu_2-A^{(n)}(\eta_2))}
{(\mu_2-A^{(n)}(\eta_2))B^{(n)}(\eta_1)-(\mu_1-A^{(n)}(\eta_1))B^{(n)}(\eta_2)},
\end{equation}
\begin{equation}
\label{g2c}
m_2=\frac {(\eta_2-\eta_1)(\mu_1-A^{(n)}(\eta_1))B^{(n)}(\eta_2)}
{(\mu_2-A^{(n)}(\eta_2))B^{(n)}(\eta_1)-(\mu_1-A^{(n)}(\eta_1))B^{(n)}(\eta_2)}.
\end{equation}
\end{subequations}
By substituting (\ref{g2}), (\ref{g1}) gives rise to infinite number ($n=0,1,...$) 
of explicit two-point BT
for the constrained flows (\ref{s4}). We now show that the two-point BTs 
(\ref{g1}) are canonical transformations and possess the spectrality property.
It is easy to check that
\begin{eqnarray}
\label{g3}
&m_2^2+(\eta_2-\eta_1)m_2-m_1m_3=0,\nonumber\\
&f_1=\frac {m_2}{m_1},\qquad f_2=\frac {m_2+\eta_2-\eta_1}{m_1}.
\end{eqnarray}
Using (\ref{g3}),  (\ref{g1b}) and (\ref{g1c})  can be rewritten as
\begin{subequations}
\label{g4}
\begin{equation}
\label{g4a}
\phi_{2j}=\frac 
1{m_1}[(\la_j-\eta_1+m_2)\phi_{1j}-\sqrt{(\la_j-\eta_1)(\la_j-\eta_2)}\bar\phi_{1j
}],
\end{equation}
\begin{equation}
\label{g4b}
\bar\phi_{2j}=\frac 
1{m_1}[\sqrt{(\la_j-\eta_1)(\la_j-\eta_2)}\phi_{1j}+(-\la_j+\eta_2+m_2)\bar\phi_{1
j}].
\end{equation}
\end{subequations}
The formulas (\ref{h8}), (\ref{h13}) and (\ref{g3}) lead to
\begin{subequations}
\label{g5}
$$[\la^2-\la(\eta_1+\eta_2)+\eta_1\eta_2]\overline A^{(n)}(\la)=
[\la^2-\la(\eta_1+\eta_2)+\eta_1\eta_2-2m_1m_3]A^{(n)}(\la)$$
\begin{equation}
\label{g5a}
-m_3[\la-\eta_1+m_2]B^{(n)}(\la)-m_1[\la-\eta_2-m_2]C^{(n)}(\la),
\end{equation}
$$[\la^2-\la(\eta_1+\eta_2)+\eta_1\eta_2]\overline 
B^{(n)}(\la)=2m_1[\la-\eta_1+m_2]A^{(n)}(\la)$$
\begin{equation}
\label{g5b}
+[\la-\eta_1+m_2]^2B^{(n)}(\la)-m_1^2C^{(n)}(\la),
\end{equation}
$$[\la^2-\la(\eta_1+\eta_2)+\eta_1\eta_2]\overline 
C^{(n)}(\la)=2m_3[\la-\eta_2-m_2]A^{(n)}(\la)$$
\begin{equation}
\label{g5c}
-m_3^2B^{(n)}(\la)+[\la-\eta_2-m_2]^2C^{(n)}(\la).
\end{equation}
\end{subequations}

(1) For the first constrained flow, the FDIHS (\ref{s11}), using (\ref{g3}) and 
(\ref{g5}), one gets
$$      m_1=\frac 14<\overline \Phi_1,\overline \Phi_1>-\frac 14<\Phi_1,\Phi_1>,$$
$$m_2=\frac 1{<\overline \Phi_1,\overline \Phi_1>-<\Phi_1,\Phi_1>}[<\La\overline 
\Phi_1,\overline \Phi_1>+<\La \Phi_1,\Phi_1>$$
\begin{equation}
\label{g6}
-\eta_2<\overline \Phi_1,\overline \Phi_1>-\eta_1<\Phi_1,\Phi_1>
-2\sum_{j=1}^{N}\sqrt{(\la_j-\eta_1)(\la_j-\eta_2)}\phi_{1j}\bar\phi_{1j}].
\end{equation}
Then substituting (\ref{g6}) into (\ref{g4}) gives rise to
\begin{equation}
\label{g7}
	\phi_{2j}=\frac {\p S^{(0)}}{\p \phi_{1j}},\qquad
\overline\phi_{2j}=-\frac {\p S^{(0)}}{\p \overline\phi_{1j}},\qquad j-1,...,N,
\end{equation}
where the generating function $S^{(0)}$ for the canonical transformation 
(\ref{g1}) is given by
$$S^{(0)}=
\frac 2{<\overline \Phi_1,\overline \Phi_1>-<\Phi_1,\Phi_1>}[<\La\overline 
\Phi_1,\overline \Phi_1>+<\La \Phi_1,\Phi_1>-\eta_2<\overline \Phi_1,\overline 
\Phi_1>$$
\begin{equation}
\label{g8}
-\eta_1<\Phi_1,\Phi_1>
-2\sum_{j=1}^{N}\sqrt{(\la_j-\eta_1)(\la_j-\eta_2)}\phi_{1j}\bar\phi_{1j}]
-\eta_1-\eta_2.
\end{equation}
Furthermore, it is found 
$$\frac {\p S^{(0)}}{\p \eta_1}=-1+
\frac 2{<\overline \Phi_1,\overline \Phi_1>-<\Phi_1,\Phi_1>}[-<\Phi_1,\Phi_1>
+2\sum_{j=1}^{N}\sqrt{\frac{(\la_j-\eta_2)}{(\la_j-\eta_1)}}\phi_{1j}\bar\phi_{1j}
]$$
\begin{subequations}
\begin{equation}
\label{g9a}
=A^{(0)}(\eta_1)+f_1B^{(0)}(\eta_1)=\mu_1,
\end{equation}
\begin{equation}
\label{g9b}
\frac {\p S^{(0)}}{\p \eta_2}
=A^{(0)}(\eta_2)+f_2B^{(0)}(\eta_2)=\mu_2.
\end{equation}
\end{subequations}
This implies that $(\eta_1,\mu_1)$ and $(\eta_2,\mu_2)$  satisfy the spectrality 
property and the separation equations given by the spectral curve (\ref{s13})
$$\mu_i^2=1+\sum_{j=1}^N \frac{P_j}{\eta_i-\lambda_j}, \qquad i=1,2.$$

(2) For the second constrained flow, the  FDIHS (\ref{s14}), formula (\ref{g5}) 
gives rise to
$$m_1=\frac 12(q-\bar q),$$
\begin{equation}
\label{g10}
m_2=
\frac 1{q+\bar q}[\frac 14<\overline \Phi_1,\overline \Phi_1>-\frac 
14<\Phi_1,\Phi_1>+\eta_1 q-\eta_2\bar q],
\end{equation}
and
$$r=\frac 4{(q-\bar q)^2}[-\frac 14<\La\overline \Phi_1,\overline \Phi_1>-\frac 
14<\La \Phi_1,\Phi_1>+\frac 14 (\eta_1+\eta_2)<\overline \Phi_1,\overline 
\Phi_1>$$
\begin{subequations}
\label{g11}
\begin{equation}
\label{g11a}
-\eta_1\eta_2\bar q+(m_2-\eta_1)^2q+\frac 
12\sum_{j=1}^{N}\sqrt{(\la_j-\eta_1)(\la_j-\eta_2)}\phi_{1j}\bar\phi_{1j}],
\end{equation}
\begin{equation}
\label{g11b}
\bar r=-2\frac {m_2}{m_1}(m_2+\eta_2-\eta_1)+r.
\end{equation}
\end{subequations}
By substitution of (\ref{g10}), (\ref{g4}) and (\ref{g11}) can be rewritten as
\begin{equation}
\label{g12}
	\phi_{2j}=\frac {\p S^{(1)}}{\p \phi_{1j}},\quad
\overline\phi_{2j}=-\frac {\p S^{(1)}}{\p \overline\phi_{1j}},\quad
r=\frac {\p S^{(1)}}{\p q},\quad
\bar r=-\frac {\p S^{(1)}}{\p \bar q},
\end{equation}
where the generating function $S^{(1)}$ for the canonical transformation 
(\ref{g1}) is given by
$$S^{(1)}=
\frac 1{q-\bar q}[<\La\overline \Phi_1,\overline 
\Phi_1>+<\La\Phi_1,\Phi_1>-2\sum_{j=1}^{N}\sqrt{(\la_j-\eta_1)(\la_j-\eta_2)}\phi_
{1j}\bar\phi_{1j}]$$
$$+\frac 1{q^2-\bar q^2}[\frac 14<\overline \Phi_1,\overline 
\Phi_1><\Phi_1,\Phi_1>-(\eta_1+\eta_2)q<\overline \Phi_1,\overline \Phi_1>
-(\eta_1+\eta_2)\bar q<\Phi_1,\Phi_1>$$
\begin{equation}
\label{g13}
-\frac 18<\overline \Phi_1,\overline \Phi_1>^2-\frac 18<\Phi_1,\Phi_1>^2
+4\eta_1\eta_2q\bar q
-2(\eta_1^2+\eta_2^2)\bar q^2]-\eta_1^2-\eta_2^2.
\end{equation}

It is easy to check the spectrality property
$$\frac {\p S^{(1)}}{\p \eta_1}=\frac 1{q^2-\bar q^2}[-q<\overline 
\Phi_1,\overline \Phi_1>-\bar q<\Phi_1,\Phi_1>-
4\eta_1\bar q^2+4\eta_2q\bar q]-2\eta_1$$
\begin{subequations}
\begin{equation}
\label{g14a}
+\frac 1{q-\bar 
q}\sum_{j=1}^N\sqrt{\frac{\la_j-\eta_2}{\la_j-\eta_1}}\phi_{1j}\overline\phi_{1j}=
-2[A^{(1)}(\eta_1)+f_1B^{(1)}(\eta_1)]=-2\mu_1,
\end{equation}
\begin{equation}
\label{g14b}
\frac {\p S^{(1)}}{\p \eta_2}
=-2[A^{(1)}(\eta_2)+f_2B^{(1)}(\eta_2)]=-2\mu_2.
\end{equation}
\end{subequations}
The points $(\eta_i,\mu_i)$ satisfy the separation equations given by the spectral 
curve (\ref{s16})
$$\mu_i^2=\eta_i^2+P_0+\sum_{j=1}^N \frac{P_j}{\eta_i-\lambda_j}, \qquad i=1,2.$$

(3) For the third constrained flow, the  FDIHS (\ref{s17}), by means of (\ref{g3}) 
and (\ref{g5}), one gets
$$m_1=\frac 12(q_1-\bar q_1),\qquad
m_3=\frac 12(q_2-\bar q_2),$$
\begin{equation}
\label{g15}
m_2=
\frac 12(\eta_1-\eta_2)+\frac 12\sqrt{(\eta_1-\eta_2)^2+(q_1-\bar q_1)(q_2-\bar 
q_2)},
\end{equation}
and
\begin{subequations}
\label{g16}
$$p_1=
\frac 2{(q_1-\bar q_1)^2}\{-\frac 12<\La\Phi_1,\Phi_1>-\frac 12<\La\overline 
\Phi_1,\overline \Phi_1>+\frac 12(\eta_1+\eta_2)<\overline \Phi_1,\overline 
\Phi_1>$$
$$-m_1(m_2-\eta_1)^2q_1q_2
+(m_2-\eta_1)^2(m_2 q_1+m_2\bar q_1+\eta_2 \bar q_1-\eta_1 q_1)$$
$$+\sum_{j=1}^{N}\sqrt{(\la_j-\eta_1)(\la_j-\eta_2)}\phi_{1j}\bar\phi_{1j}$$
$$+\frac {\eta_1\eta_2-(m_2-\eta_1)^2}{\sqrt{(\eta_1-\eta_2)^2+(q_1-\bar 
q_1)(q_2-\bar q_2)}}[
-\frac 12<\overline \Phi_1,\overline \Phi_1>+\frac 12<\Phi_1,\Phi_1>
$$
\begin{equation}
\label{g16a}
-\frac 14 q_2(q_1^2-\bar q_1^2)+(m_2-\eta_1)^2(q_1+2\bar 
q_1)+2(m_2-\eta_1)(\eta_1+\eta_2)\bar q_1
+\eta_1\eta_2\bar q_1]\},
\end{equation}
\begin{equation}
\label{g16b}
\bar p_1=p_1+\frac 12(m_2+\eta_2)(q_2+\bar q_2)-\frac 12(\eta_1+\eta_2)\bar q_2,
\end{equation}
$$\bar p_2=\frac {1}{2\sqrt{(\eta_1-\eta_2)^2+(q_1-\bar q_1)(q_2-\bar q_2)}}[
-\frac 12<\overline \Phi_1,\overline \Phi_1>+\frac 12<\Phi_1,\Phi_1>$$
\begin{equation}
\label{g16c}
-\frac 14q_2( q_1^2-\bar q_1^2)+(m_2-\eta_1)^2(q_1+2\bar q_1)
+2(m_2-\eta_1)(\eta_1+\eta_2)\bar q_1+\eta_1\eta_2\bar q_1],
\end{equation}
\begin{equation}
\label{g16d}
p_2=\bar p_2+\frac 12(\eta_1+\eta_2)q_1-\frac 12(m_2+\eta_2)(q_1+\bar q_1).
\end{equation}
\end{subequations}
By inserting (\ref{g15}) into (\ref{g4}) and (\ref{g16}), a straightforward 
calculation leads to
$$\phi_{2j}=\frac {\p S^{(2)}}{\p \phi_{1j}},\qquad
\overline\phi_{2j}=-\frac {\p S^{(2)}}{\p \overline\phi_{1j}},\qquad
p_1=\frac {\p S^{(2)}}{\p q_1},$$
\begin{equation}
\label{g17}
p_2=\frac {\p S^{(2)}}{\p q_2},\qquad
\bar p_1=-\frac {\p S^{(2)}}{\p \bar q_1},\qquad
\bar p_2=-\frac {\p S^{(2)}}{\p \bar q_2},
\end{equation}
where the generating function $S^{(2)}$ for the canonical transformation  
(\ref{g1}) is given by
$$S^{(2)}=
\frac 1{q_1-\bar q_1}[<\La\overline \Phi_1,\overline 
\Phi_1>+<\La\Phi_1,\Phi_1>-\frac 12(\eta_1+\eta_2)(<\overline \Phi_1,\overline 
\Phi_1>+<\Phi_1,\Phi_1>)$$
$$+\frac 23(m_2-\eta_1)^3(q_1+2\bar q_1)
+2(m_2-\eta_1)^2(\eta_1+\eta_2)\bar q_1-(\eta_1+\eta_2)\eta_1\eta_2\bar q_1$$
$$+\frac 12 \sqrt{(\eta_1-\eta_2)^2+(q_1-\bar q_1)(q_2-\bar q_2)}
(<\Phi_1,\Phi_1>-<\overline \Phi_1,\overline \Phi_1>
+2\eta_1\eta_2\bar q_1)$$
$$-2\sum_{j=1}^{N}\sqrt{(\la_j-\eta_1)(\la_j-\eta_2)}\phi_{1j}\bar\phi_{1j}]+\frac 
14(\eta_1+\eta_2)(q_1-\bar q_1) q_2$$
\begin{equation}
\label{g17}
-\frac 14 \sqrt{(\eta_1-\eta_2)^2+(q_1-\bar q_1)(q_2-\bar q_2)}q_2(q_1+\bar q_1)
+\frac 13(\eta_1^3+\eta_2^3).
\end{equation}
By a straightforward calculation, we can show  the spectrality property
\begin{equation}
\label{g18}
\frac {\p S^{(2)}}{\p \eta_1}=
A^{(2)}(\eta_1)+f_1B^{(2)}(\eta_1)=\mu_1,\qquad
\frac {\p S^{(2)}}{\p \eta_2}=
A^{(2)}(\eta_2)+f_2B^{(2)}(\eta_2)=\mu_2.
\end{equation}
The points $(\eta_i,\mu_i)$ satisfy the separation equations given by the spectral 
curve (\ref{s19})
$$\mu_i^2=\eta_i^4+P_0\eta_i+P_{N+1}+\sum_{j=1}^N 
\frac{P_j}{\eta_i-\lambda_j},\qquad i=1,2.$$

\section{m-times repeated two-point DTs for high-order constrained flows of the 
AKNS hierarchy}
\setcounter{equation}{0}
\hskip\parindent

Assume that $(\psi_1(x,\eta_i), \psi_2(x,\eta_i))^T, i=1,...,2m,$ be 
solutions of (\ref{s7}) and (\ref{s8}) with $\la=\eta_i, \mu=\mu_i,
i=1,2,...,2m, \eta_i\neq\la_j.$ We use $q[l], r[l], \phi_{1j}[l],
\phi_{2j}[l]$ to denote the action of l-times repeated two-point
DTs of (\ref{g1}) on the initial solution $q[0], r[0], \phi_{1j}[0], 
\phi_{2j}[0]$. We have according to (\ref{g1})
\begin{subequations}
\label{f1}
\begin{equation}
\label{f1a}
q[l+1]=q[l]-2m_1[l], \qquad r[l+1]=r[l]-2m_3[l],
\end{equation}
\begin{equation}
\label{f1b}
	\phi_{1j}[l+1]=\frac 
1{\sqrt{(\la_j-\eta_{2l+1})(\la_j-\eta_{2l+2})}}[(\la_j-\eta_{2l+1}+m_2[l])\phi_{1
j}[l]-m_1[l]\phi_{2j}[l]],
\end{equation}
\begin{equation}
\label{f1c}
\phi_{2j}[l+1]=\frac 
1{\sqrt{(\la_j-\eta_{2l+1})(\la_j-\eta_{2l+2})}}[m_3[l]\phi_{1j}[l]+(\la_j-\eta_{2
l+2}-m_2[l])\phi_{2j}[l]].
\end{equation}
\end{subequations}
We denote
$$G_m=\left( \begin{array}{ccccc} 
\eta_1^m\psi_1(\eta_1)&\eta_2^m\psi_1(\eta_2)&\eta_3^m\psi_1(\eta_3)&......& 
\eta_{2m}^m\psi_1(\eta_{2m}) 
\\\eta_1^{m-1}\psi_1(\eta_1)&\eta_2^{m-1}\psi_1(\eta_2)&\eta_3^{m-1}\psi_1(\eta_3)
&......&\eta_{2m}^{m-1}\psi_1(\eta_{2m})\\......&......&......&......&......\\
\psi_1(\eta_1)&\psi_1(\eta_2)&\psi_1(\eta_3)&......& \psi_1(\eta_{2m}) 
\\\eta_1^{m-2}\psi_2(\eta_1)&\eta_2^{m-2}\psi_2(\eta_2)&\eta_3^{m-2}\psi_2(\eta_3)
&......&\eta_{2m}^{m-2}\psi_2(\eta_{2m})\\......&......&......&......&......\\
\psi_2(\eta_1)&\psi_2(\eta_2)&\psi_2(\eta_3)&......& \psi_2(\eta_{2m}) 
 \end{array} \right),$$

$$\triangle_m=\left( \begin{array}{ccccc} 
\eta_1^{m-1}\psi_1(\eta_1)&\eta_2^{m-1}\psi_1(\eta_2)&\eta_3^{m-1}\psi_1(\eta_3)&.
.....&\eta_{2m}^{m-1}\psi_1(\eta_{2m})
 
\\\eta_1^{m-2}\psi_1(\eta_1)&\eta_2^{m-2}\psi_1(\eta_2)&\eta_3^{m-2}\psi_1(\eta_3)
&......&\eta_{2m}^{m-2}\psi_1(\eta_{2m})\\......&......&......&......&......\\
\psi_1(\eta_1)&\psi_1(\eta_2)&\psi_1(\eta_3)&......& \psi_1(\eta_{2m}) 
\\\eta_1^{m-1}\psi_2(\eta_1)&\eta_2^{m-1}\psi_2(\eta_2)&\eta_3^{m-1}\psi_2(\eta_3)
&......&\eta_{2m}^{m-1}\psi_2(\eta_{2m})\\......&......&......&......&......\\
\psi_2(\eta_1)&\psi_2(\eta_2)&\psi_2(\eta_3)&......& \psi_2(\eta_{2m}) 
 \end{array} \right),$$

\ \
$$W_m(j)=\left( \begin{array}{ccccc} 
\la_j^m\phi_{1j}[0]&\eta_1^m\psi_1(\eta_1)&\eta_2^m\psi_1(\eta_2)&......& 
\eta_{2m}^m\psi_1(\eta_{2m}) 
\\\la_j^{m-1}\phi_{1j}[0]&\eta_1^{m-1}\psi_1(\eta_1)&\eta_2^{m-1}\psi_1(\eta_2)&..
....&\eta_{2m}^{m-1}\psi_1(\eta_{2m})\\......&......&......&......&......\\
\phi_{1j}[0]&\psi_1(\eta_1)&\psi_1(\eta_2)&......& \psi_1(\eta_{2m}) 
\\\la_j^{m-1}\phi_{2j}[0]&\eta_1^{m-1}\psi_2(\eta_1)&\eta_2^{m-1}\psi_2(\eta_2)&..
....&\eta_{2m}^{m-1}\psi_2(\eta_{2m})\\......&......&......&......&......\\
\phi_{2j}[0]&\psi_2(\eta_1)&\psi_2(\eta_2)&......& \psi_2(\eta_{2m}) 
 \end{array} \right).$$
Then $m$-times repeated DTs for the constrained flows (\ref{s4}) are given by
\begin{subequations}
\begin{equation}
\label{f2a}
q[m]=q[0]-2\frac {G_m}{\triangle_m^*}, \qquad 
r[m]=r[0]-2\frac {G_m^*}{\triangle_m^*},
\end{equation}
\begin{equation}
\label{f2b}
	\phi_{1j}[m]=\frac 1{\sqrt{\prod_{i=1}^{2m}(\la_j-\eta_i)}}\frac 
{W_m(j)}{\triangle_m}, \qquad 
\phi_{2j}[m]=\frac 1{\sqrt{\prod_{i=1}^{2m}(\la_j-\eta_i)}}\frac 
{W^*_m(j)}{\triangle_m^*},
\end{equation}
\end{subequations}
where $G_m^*$, $W^*_m(j)$ and $\triangle_m^*$ are obtained by interchanging
$\psi_1(x,\eta_i)$ and $\psi_2(x,\eta_i), i=1,...,2m, \phi_{1j}(\la_j)$ and 
$\phi_{2j}(\la_j)$ in $G_m, W_m(j)$ and  $\triangle_M$, respectively. Formula 
(\ref{f2a}) was shown in \cite {1,12}, (\ref{f2b}) can be obtained in the same way 
by using the formulas for Vandermonde-like determinants $G_m^*$, $W^*_m(j)$ and 
$\triangle_m^*$ given in \cite{17}.

\section{Conclusion}
\setcounter{equation}{0}
\hskip\parindent
Some methods were presented to construct the  B\"{a}cklund transformations  with 
the properties described above in  \cite{2,3,4,5,6} for few examples. In this 
paper we propose a way to construct infinite number of explicit one- and two-point 
B\"{a}cklund transformations for high-order constrained flows of soliton hierarchy 
by means of the Darboux transformations for the constrained flows by using the 
high-order constrained flows of the AKNS hierarchy as model. By constructing the 
generating functions, it is shown that these BTs are canonical transformations 
including B\"{a}cklund parameter $\eta$  and  a spectrality property holds with 
respect to the B\"{a}cklund parameter $\eta$ and the conjugate variable $\mu$. The 
pair $(\eta,\mu)$ lies on the spectral curve and satisfies the separation 
equation. Also we 
present the formula for m-times repeated Darboux transformations for
the high-order constrained flows of the AKNS hierarchy.

The method prososed in this paper can be applied to the high-order binary
constrained flows of AKNS hierarchy in \cite{ma} to find new explicit BT's
with canonicity and spectrality.

\ \
\section*{ Acknowledgment }
The work described in this paper was supported by a grant from CityU (project No. 
7001072) and 
the Special Funds for Chinese Major Basic Research Project "Nonlinear Science".


\end{document}